# Optimal Scheduling of Integrated Demand Response-Enabled Integrated Energy Systems with Uncertain Renewable Generations: A Stackelberg Game Approach

Yang Li [a,*], Chunling Wang [a], Guoqing Li [a], Chen Chen [b]

[a] School of Electrical Engineering, Northeast Electric Power University, Jilin 132012, China
[b] School of Electrical Engineering, Xi'an Jiaotong University, Xi'an 710049, China

**Abstract**: In order to balance the interests of integrated energy operator (IEO) and users, a novel Stackelberg game-based optimization framework is proposed for the optimal scheduling of integrated demand response (IDR)-enabled integrated energy systems with uncertain renewable generations, where the IEO acts as the leader who pursues the maximization of his profits by setting energy prices, while the users are the follower who adjusts energy consumption plans to minimize their energy costs. Taking into account the inherent uncertainty of renewable generations, the probabilistic spinning reserve is written in the form of a chance constraint; in addition, a district heating network model is built considering the characteristics of time delay and thermal attenuation by fully exploiting its potential, and the flexible thermal comfort requirements of users in IDR are considered by introducing a predicted mean vote (PMV) index. To solve the raised model, sequence operation theory is introduced to convert the chance constraint into its deterministic equivalent form, and thereby, the leader-follower Stackelberg game is tackled into a mixed-integer quadratic programming formulation through Karush-Kuhn-Tucker optimality conditions and is finally solved by the CPLEX optimizer. The results of two case studies demonstrate that the proposed Stackelberg game-based approach manages to achieve the Stackelberg equilibrium between IEO and users by the coordination of renewable generations and IDR. Furthermore, the study on a real integrated energy system in China verifies the applicability of the proposed approach for real-world applications.

## NOMENCLATURE

**Abbreviations**

| | |
|---|---|
| IES | Integrated energy system |
| IEO | Integrated energy operator |
| IDR | Integrated demand response |
| CHP | Combined heat and power |
| DHN | District heating network |
| SOT | Sequence operation theory |
| KKT | Karush-Kuhn-Tucker |
| PV | Photovoltaic |
| WT | Wind turbine |
| TP | Thermal power |
| BESS | Battery energy storage system |
| HE | Heat exchanger |
| PMV | Predicted mean vote |
| RTP | Real-time price |
| DHS | District heating system |
| PDF | Probability density function |

**Sets and Parameters**

| | |
|---|---|
| $n$ | Set of indices of TP units |
| $N$ | Set of indices of CHP units |
| $T$ | Set of indices of scheduling periods |
| $\Delta t^d$ | Flow time of water (°C) |
| $P_t^{FL}$ | Fixed electric load in period $t$ (MW) |
| $H_t^{OL}$ | Original heating load in period $t$ (MW) |
| $\omega_i, \varphi_i, \kappa$ | Reserve cost coefficients of TP unit $i$, CHP unit $i$ and BESS ($/MW) |
| $g_1, g_2$ | Discharge and charge cost coefficients of BESS ($/MW) |
| $T_t^{in}$ | Indoor temperature (°C) |
| $T_{\min}^{sw}, T_{\max}^{sw}$ | Minimum and maximum temperature of the supply water (°C) |
| $T_{\min}^{rw}, T_{\max}^{rw}$ | Minimum and maximum temperature of the return water (°C) |
| $\mu_{\min}, \mu_{\max}$ | Minimum and maximum electricity price ($/MWh) |
| $\gamma_{\min}, \gamma_{\max}$ | Minimum and maximum thermal price ($/MWh) |
| $\mu_{av}, \gamma_{av}$ | Average electricity and thermal price ($/MWh) |

**Variables**

| | |
|---|---|
| $P_t^{PV}$ | PV power output in period $t$ (MW) |
| $P_t^{WT}$ | WT power output in period $t$ (MW) |
| $T_t^{sw}$ | Water supply temperature in period $t$ (°C) |
| $T_t^{rw}$ | Water return temperature in period $t$ (°C) |
| $P_t^{SL}$ | Time-shiftable electric load in period $t$ (MW) |
| $H_t^{CL}$ | Cuttable heating load in period $t$ (MW) |
| $\mu_t$ | Real-time electricity price in period $t$ ($/MWh) |
| $\gamma_t$ | Real-time thermal price in period $t$ ($/MWh) |
| $P_{i,t}^{TP}$ | Electric power of TP unit $i$ in period $t$ (MW) |
| $P_{i,t}^{CHP}$ | Electric power of CHP unit $i$ in period $t$ (MW) |
| $H_{i,t}^{CHP}$ | Heat power of CHP unit $i$ in period $t$ (MW) |
| $R_{i,t}^{TP}, R_{i,t}^{CHP}$ | Reserve capacities of TP unit $i$, CHP unit $i$ in period $t$ (MW) |
| $P_t^{CH}, P_t^{DH}$ | Charge and discharge power of BESS (MW) |
| $R_t^{BESS}$ | Reserve capacity of the BESS in period $t$ (MW) |
| $P^R$ | Consumption power of renewables (MW) |

* Corresponding author. E-mail address: liyang@neepu.edu.cn (Y. Li).

# 1 Introduction

With growing concerns on environmental pollution and energy crisis together with the low-carbon transition requirements, the development and utilization of renewable energy resources have become an indispensable choice for ensuring clean and sustainable energy supply [1]. However, the inherent uncertainty of renewable generations and the traditional "heat-set" operating mode of combined heat and power (CHP) units greatly limit the level of renewable energy consumption. In recent years, integrated energy systems (IESs) have received increasing attention from academia and industry since they break the barriers of conventional physical isolations among various energy systems and conform to the trend of renewable energy development [2], [3]. For an IES, different forms of energies including renewables are increasingly coupled in production, transmission and consumption. In term of how to handle the uncertainty of renewable generations, robust programming which sacrifices economy in exchange for reliability is conservative, while chance constraint programming, as an important stochastic programming method, can achieve the trade-off between reliability and economy. As an extension of conventional demand response (DR), integrated demand response (IDR), which was originally proposed by Aras Sheikhi and Shahab Bahrami in 2015 [4], utilizes the coupling and complementary relationships of multiple energies to optimize the operations of flexible loads, energy storage and energy conversion equipment on the demand side and eventually improves the system operation flexibility and energy utilization efficiency [5]. Meanwhile, the optimal operation of an IES is closely related to the coordination of different stakeholders and game theory is proven as an effective tool to trade-off their profits [6]. Therefore, how to balance the interests of multiple stakeholders using game theory in the optimal scheduling of an IES with renewables considering IDR is an urgent and challenging problem.

## 1.1 Literature review

Up to now, there have been some important pioneering studies on the scheduling of IESs. (1) In terms of renewable consumption, reference [7] uses electric boilers to decouple the electro-thermal coupling relationship of CHP units, which can improve the operational flexibility of CHP units and accordingly reduce the wind energy curtailments. The approach in [8] consumes the curtailments of wind power by enhancing the system operation flexibility resulting from utilizing hydrogen energy. Reference [9] leverages electric vehicles to increase the electric load and thereby promote renewable integration. (2) Regarding DR, an optimal operation model of an IES considering price-based demand response is proposed to improve the energy utilization efficiency in [10]. Reference [11] uses DR strategies such as energy substitution and load shifting to provide more balancing powers. In [12], a DR program is presented to enhance the potential of peak-load shifting. Reference [13] coordinates the price-based DR and flexibility measures to reduce the operating costs of users. An electric-gas IES model with an incentive-based DR strategy is put forward to regulate peak-valley loads in [14]. (3) Regarding game optimization, a Stackelberg game between distributed energy stations and energy users is proposed in [15] to analyze multiple energies transaction. In [16], a non-cooperative game among integrated energy service providers is presented to maximize their profits. Reference [17] proposes a non-cooperative game-theoretic model between demand response aggregators who compete to sell energy stored in storage equipment to obtain payoff.

The aforementioned references effectively promote the consumption of renewable energies or successfully apply DR to system scheduling. However, there are still some research gaps in IES scheduling. (1) The works in [7,8] have merely focused on single-type renewable integration; while the studies in [7-9] all neglect how to deal with renewable generation uncertainties which play an important role in a scheduling strategy. (2) In addition, although many recent studies [10-13] have integrated multiple energy forms in DR, the controllable heating load in these studies is always simply tackled as a proportion of the total heating demand or a function of indoor temperature with a fixed range, which fails to adapt different thermal comfort requirements of users in various periods; and the characteristics of district heating networks (DHNs) are rarely taken into account so that the potential of DHNs cannot be fully exploited. (3) More importantly, the game optimizations in recent references [15-17] haven't considered renewable generations in IES scheduling. To the authors' best knowledge, so far there is little or no published research on IES scheduling that coordinates IDR and multiple renewable generations to balance integrated energy operator (IEO) and users using game theory.

## 1.2 Contributions and paper organization

In order to address the above problems, this paper proposes a Stackelberg game-based optimization framework for the optimal scheduling of an IDR-enabled IES with uncertain renewable generations. The main contributions are as follows:

(1) To balance the interests of IEO and users, a novel Stackelberg game-based optimization framework coordinating multiple uncertain renewable generations and price-based IDR is proposed.
(2) Taking into account the characteristics of time delay and thermal attenuation of DHNs, a sophisticated DHN model is established and integrated into the IES scheduling, and the flexible thermal comfort requirements of users in IDR are considered by introducing a predicted mean vote (PMV) index.
(3) To solve the raised scheduling model, sequence operation theory (SOT) is introduced to convert the chance constraint of a spinning reserve into its deterministic equivalence form, which is beneficial to stably and fast find the Stackelberg equilibrium solution. Furthermore, the existence and uniqueness of Stackelberg equilibrium are proved in detail.
(4) The proposed method has been successfully applied to two IESs, including a real-world system in China. The findings demonstrate the approach manages to obtain the Stackelberg equilibrium solution efficiently. Besides, the approach is able to promote renewable consumption while maintaining consumers' thermal comfort within an acceptable range.

The remaining sections of this paper are organized as follows. The physical modeling and the Stackelberg game-based scheduling model of the IES are respectively given in Sections 2 and 3. The solution methodology is provided in Section 4. Section 5 presents case studies, and finally, the main conclusions are drawn in Section 6.

## 2  Physical modeling of IES

The schematic diagram of the IES studied in this work is shown in Fig. 1. The system composed of traditional generations, i.e. CHP units and thermal power (TP) units, renewable generations, i.e. photovoltaic (PV) panels and wind turbines (WTs), a battery energy storage system (BESS), heat exchangers (HEs) and controllable loads is a general IES with typical components.

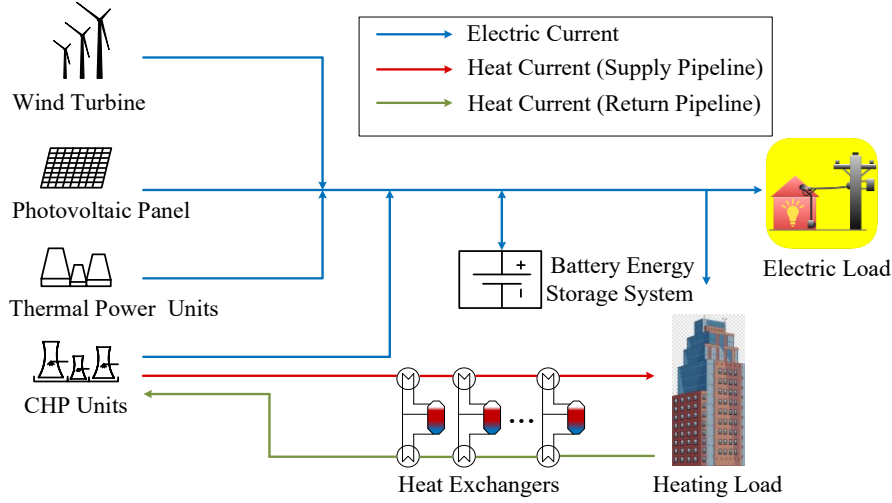

Fig. 1. Schematic diagram of the IES

### 2.1  Stochastic photovoltaic unit model

Existing studies show that solar irradiance $r$ within a certain time period approximately obeys the Beta distribution [18]:

$$f_r(r) = \frac{\Gamma(\lambda_1) + \Gamma(\lambda_2)}{\Gamma(\lambda_1)\Gamma(\lambda_2)} \left(\frac{r}{r_{max}}\right)^{\lambda_1 - 1} \left(1 - \frac{r}{r_{max}}\right)^{\lambda_2 - 1} \quad (1)$$

where $\Gamma(\cdot)$ is the Gamma function, $\lambda_1$ and $\lambda_2$ are the shape parameters. The PV power output $P^{PV}$ is

$$P^{PV} = rA^{PV}\eta^{PV} \quad (2)$$

where $A^{PV}$ is the PV radiation area, and $\eta^{PV}$ denotes the conversion efficiency. According to the Eqs. (1) and (2), the probability density function (PDF) of $P^{PV}$ can be obtained as

$$f_P(P^{PV}) = \frac{\Gamma(\lambda_1) + \Gamma(\lambda_2)}{\Gamma(\lambda_1)\Gamma(\lambda_2)} \left(\frac{P^{PV}}{P^{PV}_{max}}\right)^{\lambda_1 - 1} \left(1 - \frac{P^{PV}}{P^{PV}_{max}}\right)^{\lambda_2 - 1} \quad (3)$$

where $P^{PV}_{max}$ is the maximum power output of a PV unit.

### 2.2  Stochastic wind turbine unit model

Weibull distribution is widely used for depicting the probability distribution of wind speeds [19], which is

$$f_v(v) = u/z(v/z)^{u-1} \exp[-(v/z)^u] \quad (4)$$

where $v$, $z$ and $u$ are the actual wind speed, the scale factor and the shape factor, respectively. The WT output $P^{WT}$ can be expressed by the following piecewise function:

$$P^{WT} = \begin{cases} 0 & v < v_{in}, v \geq v_{out} \\ [(v - v_{in})/(v_e - v_{in})]P_e & v_{in} \leq v < v_e \\ P_e & v_e \leq v < v_{out} \end{cases} \quad (5)$$

where $P_e$ and $v_e$ are the WT rated power output and the rated wind speed, respectively; $v_{in}$ and $v_{out}$ denote cut-in and cut-out wind speed. Combining (4) and (5), the PDF of $P^{WT}$ can be given as

$$f_W(P^{WT}) = \begin{cases} 0, & P^{WT} < 0 \\ 1 - \exp\{-[(1 + hP^{WT}/P_e)v_{in}/z]^u\} & 0 \leq P^{WT} < P_e \\ \quad + \exp[-(v_{out}/z)^u & \\ 1, & P^{WT} \geq P_e \end{cases} \quad (6)$$

where $h = (v_e/v_{in}) - 1$.

### 2.3 District heating network model

In general, a DHN consists of a primary heating network and a secondary heating network. During the heating transmission process, the characteristics of DHN mainly include two aspects: thermal attenuation resulting from the thermal radiation and time delay due to the flow speed limitation [20]. Considering the scale of the secondary heating network is much smaller than the primary heating network, the secondary heating network is not considered in this study, and the demands of heat users is approximately equivalent to the total heating loads at HEs.

The quality-regulation mode of DHNs is here adopted [20]. The thermal energy $H_t$ transported in a pipeline in period $t$ is

$$H_t = c^w G(T_t^{sw} - T_t^{rw}) \tag{7}$$

where $c^w$ denotes the specific heat capacity of water; $G$ is the water mass flow; $T_t^{sw}$ and $T_t^{rw}$ are the water supply and return temperature in the pipeline, respectively.

According to the basic theorem of steady-state heat transfer, the heat loss $\Delta H_t$ of the pipeline with length $L$ is given as [21]:

$$\Delta H_t = 2\pi \frac{T_t^{sw} - T_t^e}{R_h} L \tag{8}$$

where $T_t^e$ denotes the average temperature outside the pipeline; $R_h$ is the average thermal resistance between the heat medium and the surrounding environment.

It's known that the delay time is approximately equal to the flow time of the water, which can be expressed as

$$\begin{aligned} \Delta t^d &= L\pi d^2 \rho^w / 4G \\ \Delta t^{sp} &= round[\Delta t^d / \Delta t] \end{aligned} \tag{9}$$

where $d$ is the inner diameter of a pipeline; $\rho^w$ is the density of water; $\Delta t^d$ is the flow time of water, $\Delta t^{sp}$ is the delay time after rounding; $\Delta t$ represents a time interval; $[\cdot]$ denotes the rounding function.

### 2.4 Load model

#### 2.4.1 Electric load model

The electric loads, including the fixed load and time-shiftable load, participate in the IES scheduling through aggregation [22], which is formulated as

$$P_t^L = P_t^{FL} + P_t^{SL} \tag{10}$$

where $P_t^L$ is the total electric load of users in period $t$; $P_t^{FL}$ and $P_t^{SL}$ are respectively the fixed load and time-shiftable load.

#### 2.4.2 Heating load model

The heating power stored in a building can be calculated according to the first-order thermodynamic model as follows [23]:

$$H_t^{OL} = \frac{[T_t^{in} - T_t^{out}] + \dfrac{K \cdot F}{c_{air} \cdot \rho_{air} \cdot V} \cdot \Delta t \cdot [T_{t-1}^{in} - T_t^{out}]}{\dfrac{1}{K \cdot F} + \dfrac{1}{c_{air} \cdot \rho_{air} \cdot V} \cdot \Delta t} \tag{11}$$

where $T_t^{in}$ and $T_t^{out}$ are the indoor and outdoor temperatures in period $t$; $K$ is the comprehensive heat transfer coefficient; $F$ and $V$ are the surface area and volume of the building; $C_{air}$ and $\rho_{air}$ are the specific heat capacity and density of the indoor air; $H_t^{OL}$ is the heating power in period $t$.

In addition, the heating loads can be cut down by a certain percentage within an acceptable thermal comfort range of users. Noted that since the initial heating load before IDR in this paper is the value at the most comfortable temperature, users prefer to reduce heating loads to achieve the goal of reducing cost. Thus, the actual heating load $H_t^L$ in period $t$ is defined as [22]:

$$H_t^L = H_t^{OL} - H_t^{CL} \tag{12}$$

where $H_t^{CL}$ denotes the users' cuttable heating load in period $t$.

To quantify the thermal comfort of users, the predicted mean vote (PMV) index is introduced as follows [24]:

$$PMV = 2.43 - \frac{3.76(T^s - T_t^{in})}{M(I_{cl} + 0.1)} \tag{13}$$

where $M$ denotes the energy metabolism rate of the human body which is related to the intensity of human activity; $I_{cl}$ is the thermal resistance of clothing; $T^s$ is the average temperature of the human skin in a comfortable state.

The PMV indexes corresponding to 7 thermal sensations of the human body are shown in Table 1. Since the users are more sensitive to heat changes during the daytime than nighttime due to frequent activities, the users' requirement of thermal comfort during the daytime is relatively higher than that at night. The PMV indexes in a scheduling cycle are shown in Fig. 2 [25].

Table 1. The states of residents under different PMV index

| PMV | -3 | -2 | -1 | 0 | +1 | +2 | +3 |
|-----|------|------|---------------|---------|----------------|------|-----|
| State | cold | cool | slightly cool | comfort | slightly warm | warm | hot |

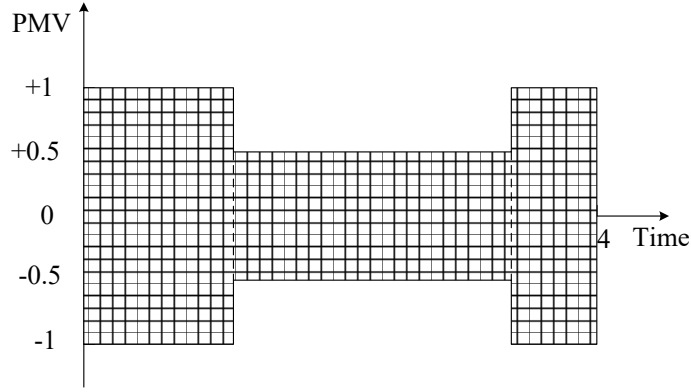

Fig. 2. The range of PMV

## 3 Proposed Stackelberg game model for IES scheduling

In the proposed game model, the IEO acts as the leader who maximizes his profits by setting prices, while the users are the followers who adjust energy consumption plans to minimize their energy costs. The IEO considered in this paper is a private entity and only transactions between users and the IEO can be achieved, however, there is no monopoly risk due to the IDR, which is an important means to prevent IEO from suppressing users.

### 3.1 Stackelberg game model formulation

During the energy transaction process, the IEO prioritizes the pricing strategy based on load demands, and the users respond to demand according to the price information. In this study, the transaction process between IEO and users is depicted in Fig. 3 and formulated as the following Stackelberg game model:

$$I = \{\{IEO, user\}; \{\mu_t, \gamma_t\}; \{P_t^{SL}, H_t^{CL}\}; \{F_1, F_2\}\} \tag{14}$$

where $\{IEO, user\}$ is the set of participants in the game; $\{\mu_t, \gamma_t\}$ and $\{P_t^{SL}, H_t^{CL}\}$ are respectively strategy sets of IEO and users in period $t$; $\{F_1, F_2\}$ is the set of profits in the game.

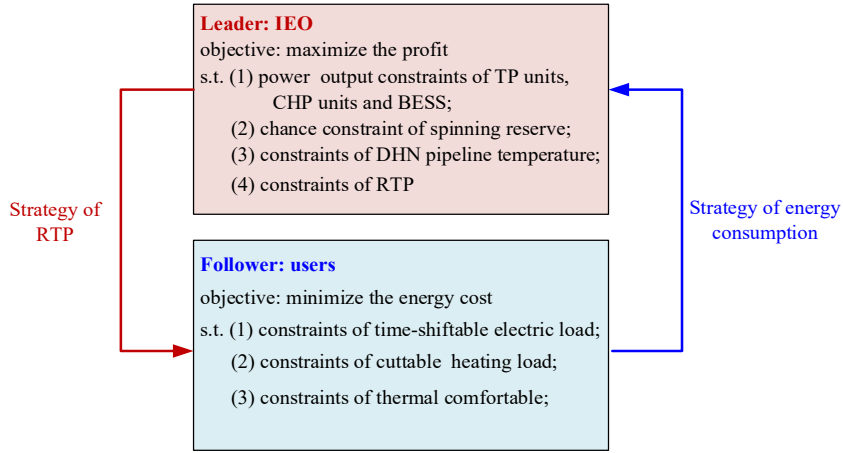

Fig. 3. Schematic diagram of the proposed Stackelberg game

### 3.2 Stackelberg game model of IEO

#### 3.2.1 Objective function

The objective function is to maximize the profits of the IEO which is the difference between the total incomes and the operating costs:

$$\max F_1 = \sum_{t=1}^{T}(\mu_t P_t^L + \gamma_t H_t^L) - C_1 - C_2 - C_3$$

$$C_1 = \sum_{t=1}^{T}\sum_{i=1}^{n}[a_i^{TP}(P_{i,t}^{TP})^2 + b_i^{TP}P_{i,t}^{TP} + c_i^{TP} + \omega_i R_{i,t}^{TP}]$$

$$C_2 = \sum_{t=1}^{T}\sum_{i=1}^{N}[a_i^{CHP}(P_{i,t}^{CHP} + c_{V,i}H_{i,t}^{CHP})^2 + b_i^{CHP}(P_{i,t}^{CHP} + c_{V,i}H_{i,t}^{CHP}) + c_i^{CHP} + \varphi_i R_{i,t}^{CHP}]$$

$$C_3 = \sum_{t=1}^{T}[g_1 P_t^{DC} + g_2 P_t^{CH} + \kappa R_t^{BESS}]$$

(15)

where $C_1$, $C_2$ represent the generation costs of the TP and CHP units, respectively; $C_3$ is the operating costs of the BESS; $n$ and $N$ are the number of TP units and CHP units; $\mu_t$ and $\gamma_t$ are the real-time electricity and thermal prices; $P_{i,t}^{TP}$ denotes the electric power of TP unit $i$; $P_{i,t}^{CHP}$ and $H_{i,t}^{CHP}$ are the electric and heat power of CHP unit $i$; $a_i^{TP}, b_i^{TP}, c_i^{TP}$ and $a_i^{CHP}, b_i^{CHP}, c_i^{CHP}$ are the fuel cost coefficients of TP and CHP unit $i$; $c_{V,i}$ is the thermal power ratio; $g_1$ and $g_2$ are the discharge and charge cost coefficients of the BESS; $P_t^{CH}$ and $P_t^{DH}$ represent the charge and discharge powers; $R_{i,t}^{TP}$, $R_{i,t}^{CHP}$, $R_t^{BESS}$ are the reserve capacities of the TP units, the CHP units and the BESS, while $\omega_i$, $\varphi_i$, $\kappa$ are the coefficients of their reserve costs.

### 3.2.2 Operational Constraints
#### 3.2.2.1 Power balance constraints
To guarantee the system secure operation during the scheduling horizon, the supply-demand balance relationship must be satisfied as follows:

$$\sum_{i=1}^{n} P_{i,t}^{TP} + \sum_{i=1}^{N} P_{i,t}^{CHP} + P^R + P_t^{DC} - P_t^{CH} = P_t^L$$

$$\sum_{i=1}^{N} H_{i,t}^{CHP} = H_{t+\Delta t^{sp}}^{L} + \Delta H_{t+\Delta t^{sp}}$$

(16)

where $P^R$ is the consumption of renewable energy sources.

#### 3.2.2.2 Ramp rate constraints
The ramp rate constraints of each TP unit and CHP unit should meet the following inequalities [23]:

$$-r_{d,i}^{TP} \leq P_{i,t}^{TP} - P_{i,t-1}^{TP} \leq r_{u,i}^{TP}$$

$$-r_{d,i}^{CHP} \leq P_{i,t}^{CHP} - P_{i,(t-1)}^{CHP} \leq r_{u,i}^{CHP}$$

(17)

where $r_{d,i}^{TP}$ and $r_{u,i}^{TP}$ are respectively the ramp-down and ramp-up rate of TP unit $i$; $r_{d,i}^{CHP}$ and $r_{u,i}^{CHP}$ are respectively the ramp-down and ramp-up rate of CHP unit $i$.

#### 3.2.2.3 Pipeline temperature constraints
The temperatures of the supply and return water are limited by the allowable temperature values of a pipeline [20]:

$$T_{\min}^{sw} \leq T_t^{sw} \leq T_{\max}^{sw}$$

$$T_{\min}^{rw} \leq T_t^{rw} \leq T_{\max}^{rw}$$

(18)

where $T_{\min}^{sw}$ and $T_{\max}^{sw}$ are the minimum and maximum temperatures of the supply water; $T_{\min}^{rw}$ and $T_{\max}^{rw}$ are the minimum and maximum temperatures of the return water.

#### 3.2.2.4 Real-time price (RTP) constraints
According to some energy policies and fairness, the real-time electricity and thermal prices should obey the following constraints [26]:

$$\mu_{\min} \leq \mu_t \leq \mu_{\max}, \gamma_{\min} \leq \gamma_t \leq \gamma_{\max}$$

$$\sum_{t=1}^{T} \mu_t = T\mu_{av}, \sum_{t=1}^{T} \gamma_t = T\gamma_{av}$$

(19)

where $\mu_{\min}$, $\mu_{\max}$ and $\mu_{av}$ are the minimum, maximum and average values of electricity prices; $\gamma_{\min}$, $\gamma_{\max}$ and $\gamma_{av}$ are the minimum, maximum and average values of thermal prices.

#### 3.2.2.5 Spinning reserve constraints
The spinning reserve constraint in the form of a chance constraint with a confidence level $\vartheta$ is formulated as [18]:

$$P_r\{\sum_{i=1}^{n} R_{i,t}^{TP} + \sum_{i=1}^{N} R_{i,t}^{CHP} + R_t^{BESS} \geq E_t - P_t^W - P_t^{PV}\} \geq \vartheta$$

(20)

where $P_r\{\}$ is the probability of an event; $E_t$ is the expected value of renewable generations. The other specific constraints of TP units, BESS and electro-thermal coupling constraints of CHP units can be found in references [18] and [23].

### 3.3 Stackelberg game model of users

#### 3.3.1 Objective function

The objective function is to minimize the energy cost, which can be expressed as follows:

$$\min F_2 = \sum_{t=1}^{T}[(\mu_t P_t^L + \gamma_t H_t^L) + \theta(H_t^{CL})^2] \quad (21)$$

where $\theta$ is the penalty factor that characterizes the users' requirements for thermal comfort; $\theta(H_t^{CL})^2$ represents the penalty cost of the users due to the reduction in comfort degree caused by cutting heating load. For the convenience of calculation, the penalty coefficients of all users can be taken as a unified value [27], and the effect of different penalty coefficients on cuttable heating loads is analyzed in this study.

### 3.3.2 Operational constraints
#### 3.3.2.1 Time-shiftable electric load constraints
The constraints of the time-shiftable load should obey the following limits:

$$\begin{aligned} \sum_{t=1}^{T} P_t^{SL} &= \alpha \sum_{t=1}^{T} P_t^L : \xi \\ P_{t,\min}^{SL} &\leq P_t^{SL} \leq P_{t,\max}^{SL} : \delta_{1,t}, \delta_{2,t} \end{aligned} \quad (22)$$

where $\alpha$ is the ratio of time-shiftable load to total electric load; $\xi$ is the Lagrange multiplier of the equality constraint; $P_{t,\min}^{SL}$ and $P_{t,\max}^{SL}$ are the minimum and maximum values of time-shiftable load; $\delta_{1,t}$ and $\delta_{2,t}$ are Lagrange multipliers of the lower and upper constraints of the time-shiftable load.

#### 3.3.2.2 Cuttable heating load constraints
To guarantee the basic need of heating load, the cuttable heating load cannot exceed its maximum value, which is given by

$$0 \leq H_t^{CL} \leq H_t^{OL} - H_{t,\min}^L : \delta_{3,t}, \delta_{4,t} \quad (23)$$

where $H_{t,\min}^L$ is the minimum value of the heating load in period $t$; $\delta_{3,t}$ and $\delta_{4,t}$ are Lagrange multipliers of the lower and upper constraints of the cuttable heating load, respectively.

#### 3.3.2.3 PMV index constraints
The minimum heating load is quantified by the following PMV index constraints:

$$\begin{cases} |PMV| \leq 0.5 & 7:00-20:00 \\ |PMV| \leq 1 & 0:00-7:00, 20:00-24:00 \end{cases} \quad (24)$$

## 4 Model solving
To solve the raised model, SOT is introduced to convert the chance constraint of spinning reserve into its deterministic equivalence form, then the leader-follower Stackelberg game is formulated as a bi-level programming problem, where the leader and follower are respectively the upper- and lower- level decision-makers; and then, the bi-level programming is tackled into a mixed-integer quadratic programming form by utilizing Karush-Kuhn-Tucker (KKT) optimality conditions.

### 4.1 Deterministic transformation of chance constraint
SOT is used to solve complex probabilistic problems by defining the operations among the sequences [28]. In this study, the SOT is adopted to handle the chance constraint of spinning reserve.

The available power outputs of WT and PV units are random variables, which can be described by probabilistic sequences. Taking the discretization step size as $q$, it can be known that the discretization sequence length $N_{at}$ of PV outputs is:

$$N_{at} = \lfloor P_{t,\max}^{PV} / q \rfloor \quad (25)$$

where $\lfloor \cdot \rfloor$ is the ceiling function; $P_{t,\max}^{PV}$ is the maximum power of PV in period $t$. Thus the PV unit has a total of $N_{at}+1$ states, where the output of state $m_a$ is $m_a q (0 \leq m_a \leq N_{at})$, and its corresponding probability is $a(m_a)$.

The probabilistic sequence $a(i_{at})$ is obtained by utilizing the PDF of the PV outputs as follows [23]:

$$a(i_{at}) = \begin{cases} \int_{0}^{q/2} f_p(P^{PV}) dP^{PV}, & i_{at} = 0 \\ \int_{i_{at}q-q/2}^{i_{at}q+q/2} f_p(P^{PV}) dP^{PV}, & i_{at} > 0 \text{ 且 } i_{at} \neq N_{at} \\ \int_{i_{at}q-q/2}^{i_{at}q} f_p(P^{PV}) dP^{PV}, & i_{at} = N_{at} \end{cases} \quad (26)$$

The processing method of the probabilistic sequence $b(i_{bt})$ of WT outputs is the same as that of PV outputs. Then the probabilistic sequence of the joint outputs $d(i_{dt})$ is calculated by the addition-type-convolution of $a(i_{at})$ and $b(i_{bt})$, i.e.

$$d(i_{dt}) = \sum_{i_{at}+i_{bt}=i_{dt}} a(i_{at})b(i_{bt}), \quad i_{dt} = 0,1,...,N_{at}+N_{bt} \quad (27)$$

The expected value of renewable generations $E_t$ can be obtained as

$$E_t = \sum_{m_{at}=0}^{N_{at}} m_{at} q a(m_{at}) + \sum_{m_{bt}=0}^{N_{bt}} m_{bt} q b(m_{bt}) \quad (28)$$

Table 2 shows the probabilistic sequence of joint renewable energy in step size $q$ and length $N_{dt}$ ($N_{dt} = N_{at} + N_{bt}$).

Table 2. Renewable power probabilistic sequence

| Power (MW)  | 0    | $q$  | ... | $m_d q$  | ... | $N_{dt} q$ |
|---|---|---|---|---|---|---|
| Probability | $d(0)$ | $d(1)$ | ... | $d(m_d)$ | ... | $dN_{dt}$ |

Equation (20) is a chance constraint which cannot be solved directly, therefore a new variable $w_m$ is introduced as

$$w_m = \begin{cases} 1, \sum_{i=1}^{n} R_{i,t}^{TP} + \sum_{i=1}^{N} R_{i,t}^{CHP} + R_t^{BESS} \geq E_t - m_{dt} q, m_{dt} = 0,1,...,N_{dt} \\ 0, \text{ otherwise} \end{cases} \quad (29)$$

And then, the chance constraint can be rewritten as

$$\sum_{m_{dt}=0}^{N_{at}+N_{bt}} w_m d(m_{dt}) \geq \vartheta \quad (30)$$

### 4.2 Conversion of KKT optimality condition

The mathematical formulation of a bi-level programming problem is generally expressed as:

$$\begin{aligned} & \text{Max } F_1(u,z) \\ & \text{s.t. } c(u,z) \geq 0, \ e(u,z) = 0, \\ & \text{Min } F_2(u,z) \\ & \text{s.t. } C(u,z) \geq 0, \ E(u,z) = 0 \end{aligned} \quad (31)$$

where $F_1(u,z)$ and $F_2(u,z)$ are the objective functions of the upper and lower level problems; $c(u,z)$, $e(u,z)$ and $C(u,z)$, $E(u,z)$ denote the inequality and equality constraints in the upper- and lower-level problems, respectively.

The lower-level problem can be transformed into KKT optimal conditions by using the following Lagrange multiplier method [27]:

$$\begin{aligned} & \text{Max } F_1(u,z) \\ & \text{s.t. } c(u,z) \geq 0, \ e(u,z) = 0, \\ & \Delta_z F_2(u,z) - \Delta_z C(u,z)^T \delta + \Delta_z E(u,z)^T \xi = 0 \\ & E(u,z) = 0 \\ & C(u,z) \geq 0, \ C(u,z) \cdot \delta = 0, \ \delta \geq 0 \Leftrightarrow 0 \leq C(u,z) \perp \delta \geq 0 \end{aligned} \quad (32)$$

And thereby, formulas (21-24) in the lower-level problem can be transformed into the following constraints:

$$\begin{aligned} & \sum_{t=1}^{T} (\mu_t - \delta_{1,t} + \delta_{2,t} + \xi) = 0 \\ & \sum_{t=1}^{T} (-\gamma_t + 2H_t^{CL} - \delta_{3,t} + \delta_{4,t}) = 0 \\ & 0 \leq P_t^{SL} - P_{t,\min}^{SL} \perp \delta_{1,t} \geq 0, \ 0 \leq P_{t,\max}^{SL} - P_t^{SL} \perp \delta_{2,t} \geq 0 \\ & 0 \leq H_t^{CL} \perp \delta_{3,t} \geq 0, \ 0 \leq H_t^{OL} - H_{t,\min}^{L} - H_t^{CL} \perp \delta_{4,t} \geq 0 \\ & \sum_{t=1}^{T} P_t^{SL} = \alpha \sum_{t=1}^{T} P_t^{L} \\ & \text{here, } H_{t,\min}^{L} = \begin{cases} H_{PMV=-0.5} & 7:00-20:00 \\ H_{PMV=-1} & 0:00-7:00, 20:00-24:00 \end{cases} \end{aligned} \quad (33)$$

where $H_{PMV=-0.5}$ and $H_{PMV=-1}$ are respectively the corresponding heating loads when the PMV indexes are taken as -0.5 and -1.

After the conversion of KKT optimality conditions, the inequality constraints in the lower problem are transformed into complementary constraints whose form is a nonlinear structure. In this regard, the Big-M method is introduced to linearize the constraints. Taking $0 \leq P_t^{SL} - P_{t,\min}^{SL} \perp \delta_{1,t} \geq 0$ as an example, it can be converted into the following linear constraint form:

$$\begin{aligned} & 0 \leq P_t^{SL} - P_{t,\min}^{SL} \leq \varpi M \\ & 0 \leq \delta_{1,t} \leq (1-\varpi) M \end{aligned} \quad (34)$$

where $\varpi$ is a 0-1 variable. The Lagrange multiplier $\delta_{1,t}$ is 0 if $\varpi$ is 1; otherwise, $P_t^{SL} - P_{t,\min}^{SL}$ is 0 if $\varpi$ is 0. This is equivalent to the original complementary constraint in (33).

## 4.3 Existence and uniqueness of equilibrium solution

### 4.3.1 Proof of existence

When the Stackelberg game meets the following conditions, there exists an equilibrium solution:
(1) The profit of IEO is a nonempty and continuous function of strategy sets of IEO and users.
(2) The cost of users is a nonempty and continuous function of strategy sets of IEO and users.
(3) The cost of users is a concave function of the strategy set of users.

Proof: It is obvious that $F_1$ and $F_2$ are both nonempty and continuous functions of $\mu_t, \gamma_t, P_t^{SL}, H_t^{CL}$. The second-order derivatives of $F_2$ with respect to $P_t^{SL}$ and $H_t^{CL}$ respectively are:

$$\frac{\partial^2 F_2}{\partial (P_t^{SL})^2} = 0$$
$$\frac{\partial^2 F_2}{\partial (H_t^{CL})^2} = 2\theta > 0 \qquad (35)$$

Hence, $F_2$ is concave in $P_t^{SL}$ and $H_t^{CL}$. Therefore, there exists an equilibrium in the Stackelberg game.

### 4.3.2 Proof of uniqueness

When the Stackelberg game meets the following conditions, there exists a unique equilibrium solution:
(1) The cost of users has a unique minimum once informed of the strategy of IEO.
(2) The profit of IEO has a unique maximum for a given strategy of the users.

Proof (1): The derivatives of $F_2$ with respect to $P_t^{SL}$ and $H_t^{CL}$ respectively are:

$$\frac{\partial F_2}{\partial P_t^{SL}} = \mu_t > 0$$
$$\frac{\partial F_2}{\partial H_t^{CL}} = -\gamma_t + 2\theta H_t^{CL} \qquad (36)$$

Let $-\gamma_t + 2\theta H_t^{CL}$ equal to 0, it can obtain that $H_t^{CL} = \gamma_t / 2\theta$, additionally, it can be known that $F_2$ is an increased function of $P_t^{SL}$ which has a lower limit value, so the following situations may happen: when $\gamma_t / 2\theta < 0$, the optimal strategy of users is $(P_{i,\min}^{SL}, 0)$; when $0 \le \gamma_t / 2\theta \le H_t^{OL} - H_{t,\min}^{L}$, the optimal strategy is $(P_{i,\min}^{SL}, \gamma_t / 2\theta)$; when $\gamma_t / 2\theta > H_t^{OL} - H_{t,\min}^{L}$, the optimal strategy is $(P_{i,\min}^{SL}, H_t^{OL} - H_{t,\min}^{L})$. So there always has only an optimal strategy no matter in which situation.

Proof (2): The derivatives of $F_1$ with respect to $\mu_t$ and $\gamma_t$ respectively are:

$$\frac{\partial F_1}{\partial \mu_t} = P_t^{FL} + P_t^{SL} > 0$$
$$\frac{\partial F_1}{\partial \gamma_t} = H_t^{OL} - H_t^{CL} > 0 \qquad (37)$$

It can be known from (37) that $F_1$ is an increased function of $\mu_t$ and $\gamma_t$, so the profit of IEO has a unique maximum in the definition domain.

## 4.4 Solution process

The solution process of the proposed approach mainly includes the following steps:
Step 1: Build the Stackelberg game model of the IEO. (Eq. (15)-Eq. (20))
Step 2: Generate the probabilistic sequences of WT outputs, PV outputs and their joint power outputs based on the SOT. (Eq. (25)-Eq. (27))
Step 3: Obtain the expected output of renewable generations. (Eq. (28))
Step 4: Transform the chance constraint into its deterministic equivalent form. (Eq. (29)-Eq. (30))
Step 5: Build the Stackelberg game model of the users. (Eq. (21)-Eq. (24))
Step 6: Formulate the Stackelberg game model as a bi-level programming problem. (Eq. (31))
Step 7: Add real-time energy prices to the lower level.
Step 8: Transform the lower-level optimization problem into KKT optimal conditions by using Lagrange multiplier method. (Eq. (32)-Eq. (33))
Step 9: Leverage the Big-M method to transform the nonlinear constraints in KKT conditions into their linear forms. (Eq. (34))
Step 10: Add KKT optimal conditions to the upper level.
Step 11: Input the initial parameters of the IES.

Step 12: Use the CPLEX optimizer to solve the model.
Step 13: Output the optimal solution of the IES scheduling.

## 5 Case study

In order to examine the effectiveness of the adopted theoretical models and the solution methodology, two cases under different conditions are tested in this section. To be specific, the simulation is firstly carried out on an IES consisting of a modified IEEE 30-bus system and two 6-bus district heating systems (DHSs), while the second case is performed on a real IES in China to further verify the applicability of the presented IES scheduling approach for real-world applications. All numerical simulations are tested on a PC platform with Intel Core dual-core CPUs (2.4 GHz) and 8GB RAM.

In this study, four modes are selected and compared in two cases to examine the performance of the proposed approach. Note that mode 3 (joint optimization) is the mode adopted in this study; in other modes (independent optimization), the electricity/thermal prices are set to be constant and proportional to the electric/heating loads, and the heating loads are set as the value in the optimal thermal comfort state of users (PMV=0) in modes 1 and 2.

Mode 1: Independent optimization of the IEO without considering the users' cost and DHN characteristics.
Mode 2: Independent optimization of the IEO with consideration of DHN characteristics.
**Mode 3: Joint optimization of IEO and users with consideration of DHN characteristics and IDR.**
Mode 4: Independent optimization of users considering DHN characteristics.

### 5.1 Case 1: Integrated IEEE 30-bus system and two 6-bus DHSs

As seen in Fig. 4, the system which is used in the first study is composed of a modified IEEE 30-bus system and two 6-bus district heating systems (DHSs), where each HE supplies 10 buildings.

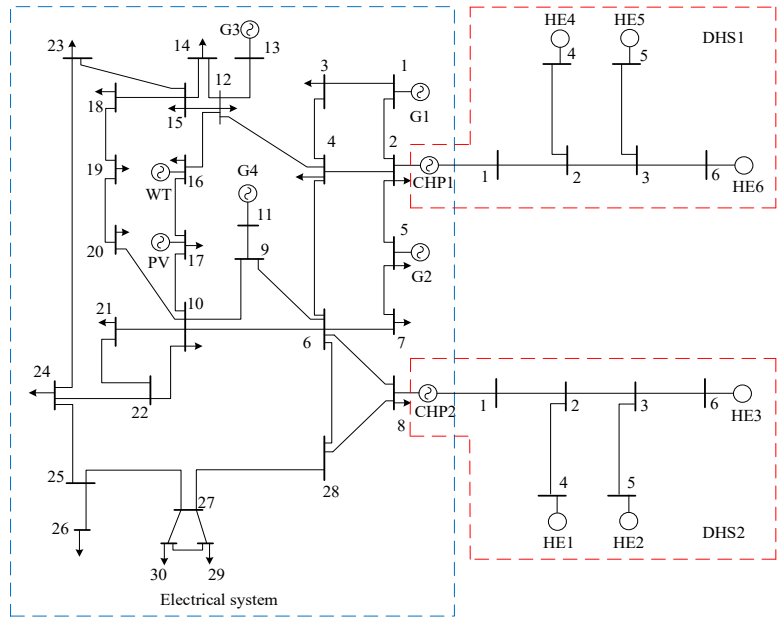

Fig. 4. Structure of the proposed IES

The main parameters of the system are as follows: Real-time price parameters are $\mu_{max}$=90$/MWh, $\mu_{min}$=40$/MWh, $\mu_{av}$=65$/MWh, $\gamma_{max}$=70$/MWh, $\gamma_{min}$=30$/MWh, $\gamma_{av}$=50$/MWh; the ratio of time-shiftable load $\alpha=10\%$; the penalty factor $\theta=0.8$; DHN parameters are $R_h = 20 \text{km}°\text{C}/\text{KW}$, $T_{min}^{sw} = 90°\text{C}$, $T_{max}^{sw} = 100°\text{C}$, $T_{min}^{rw} = 35°\text{C}$, $T_{max}^{rw} = 60°\text{C}$, $T_t^e = 0°\text{C}$; PMV parameters are $M = 80\text{M/m}^2$, $I_{cl} = 0.261\text{m}^2°\text{C/W}$, $T^s = 33.5°\text{C}$ [25]; the step size $q = 2.5\text{MW}$. The parameters of supply pipelines are shown in Table 3, and other parameters can be found in [18] and [23]. Note that the whole scheduling cycle and each scheduling period are respectively taken as 24 hours and 1 hour.

Table 3. Parameters of supply pipelines

| Pipeline | $L$ (km) | $d$ (m) | $G$ (kg/s) |
|---|---|---|---|
| DHS1:1-4 | 6.0 | 0.6 | 300 |
| DHS1:1-5 | 6.5 | 0.7 | 350 |
| DHS1:1-6 | 7.5 | 0.6 | 300 |
| DHS2:1-4 | 7.5 | 0.5 | 270 |
| DHS2:1-5 | 7.0 | 0.6 | 300 |
| DHS2:1-6 | 7.5 | 0.7 | 350 |

In this work, a typical winter day is chosen for subsequent analysis. Fig. 5 shows the outdoor temperature, the predicted value of electric load and the expected outputs of renewable generations in a whole scheduling period.

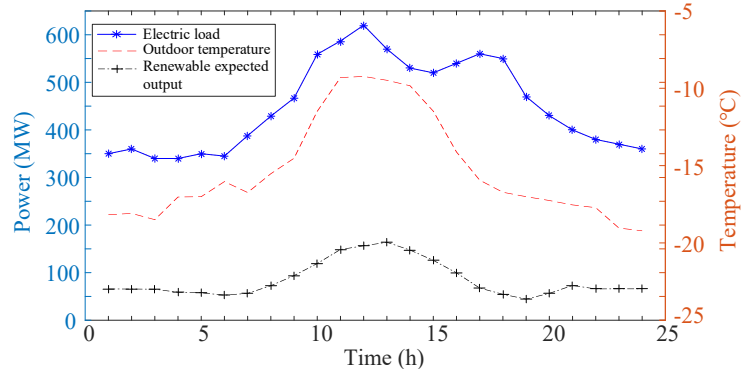

Fig. 5. Expected power of joint renewable outputs, forecasts of electric load and outdoor temperature

**5.1.1 Analysis of renewable energy consumption**

To properly evaluate the effectiveness of the proposed method on renewable consumptions, the amount of absorbed renewable energies under modes 1-3 at the 95% confidence level are analyzed with the results shown in Fig. 6 and Table 4.

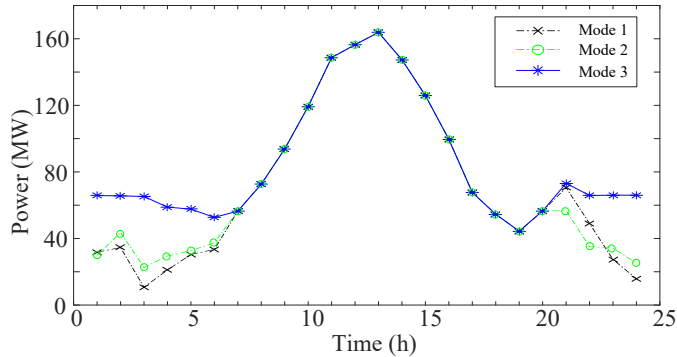

Fig. 6. Renewable consumptions under different modes

Table 4. Amounts of absorbed renewable energies under different modes

| Mode | Amounts of absorbed renewables (MW) | Excepted outputs (MW) |
| --- | --- | --- |
| Mode 1 | 1732.017 | |
| Mode 2 | 1754.318 | **2044.289** |
| **Mode 3** | **2044.289** | |

It can be seen from Fig. 6 and Table 4 that the amount of renewable consumptions from modes 1 to 3 gradually increases, which demonstrates consideration of DHN characteristics and IDR can promote renewable consumptions. Especially, mode 3 can completely absorb the renewable power outputs in this case.

**5.1.2 Analysis of the characteristics of district heating network**

The heating loads under modes 1 and 2 are compared to analyze the effects of the DHN characteristics, as shown in Fig. 7.

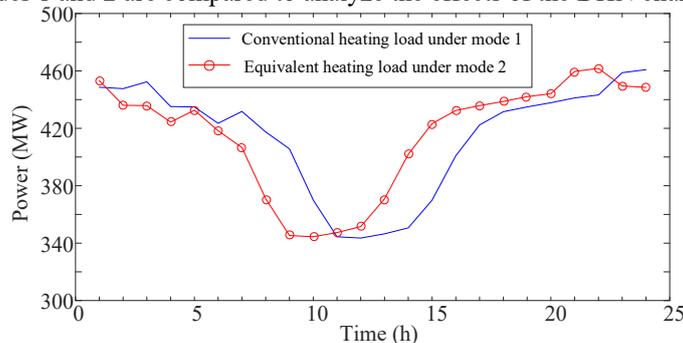

Fig. 7. Powers of the heating loads

Fig. 7 illustrates that the DHN characteristics enable the equivalent heating load to have a certain time-shifting effect. This results in the decreasement of the time overlap between the periods of heavy heating loads and high wind power generation, and thereby promoting renewable energy consumption.

The average delay time and thermal attenuation of each pipeline are shown in Table 5.

Table 5. Delay time and thermal attenuation of pipelines

| Pipeline | DHS1: 1-4 | DHS1: 1-5 | DHS1: 1-6 | DHS2: 1-4 | DHS2: 1-5 | DHS2: 1-6 |
|---|---|---|---|---|---|---|
| Delay time (h) | 1.57 | 1.98 | 1.96 | 1.51 | 1.83 | 2.29 |
| Thermal attenuation (MW) | 0.17 | 0.16 | 0.18 | 0.18 | 0.16 | 0.17 |

Table 5 shows that due to the small scale of the system, the thermal attenuation of each pipeline is relatively small, which has no significant impact on the scheduling results.

### 5.1.3 Analysis of integrated demand response

The change of electric load due to the consideration of IDR is shown in Fig. 8.

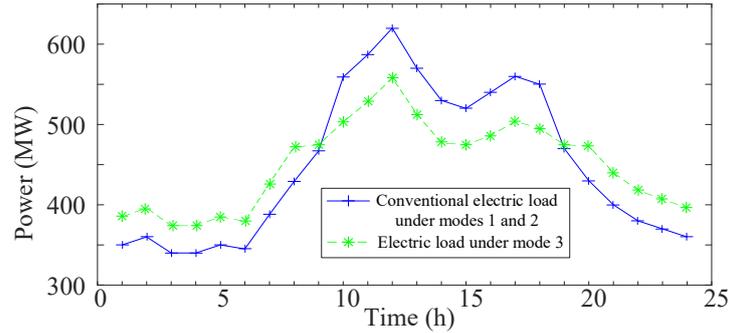

Fig. 8. Power of the electric load

As shown in Fig. 8, the users increase the electric demand during the off-peak periods and cut down the electric load during the peak-load periods under mode 3. As a result, the consumption of renewable curtailments has been significantly improved at night. This illustrates that the coordination of IDR and renewable generations is able to achieve an increase in the consumption of renewable energies.

The changes of time-shiftable electric load together with real-time electricity prices under mode 3 are shown in Fig. 9.

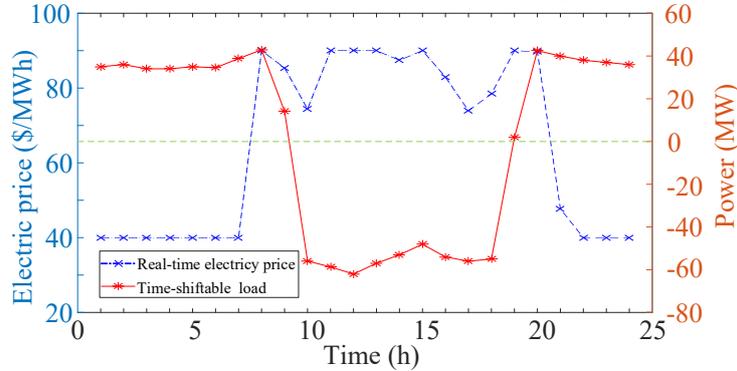

Fig. 9. Real-time electricity price and time-shiftable electric load

As shown in Fig. 9, the users shift electric demand from peak periods to off-peak periods under RTP. It suggests that the proposed real-time pricing mechanism is able to take the benefit of users into account and guide users to actively participate in IDR to decrease their energy costs.

The change of heating loads with consideration of IDR is shown in Fig. 10.

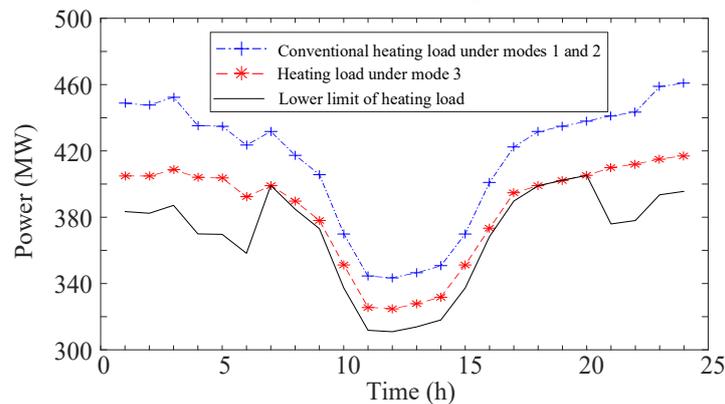

Fig. 10. Power of the heating loads

As shown in Fig. 10, the heating loads under mode 3 are reduced to varying degrees in different periods, but none of them exceed the lower limit. This shows that the users can reduce their energy cost at the expense of sacrificing some thermal comforts within an acceptable range; and that IDR can decrease the heating demand by exploiting the flexible demand elasticity on the heating load side, which promotes renewable consumption due to the reduced electric outputs of "heat-set" CHP units.

The changes of cuttable heating load in accordance with thermal price under mode 3 are shown in Fig. 11.

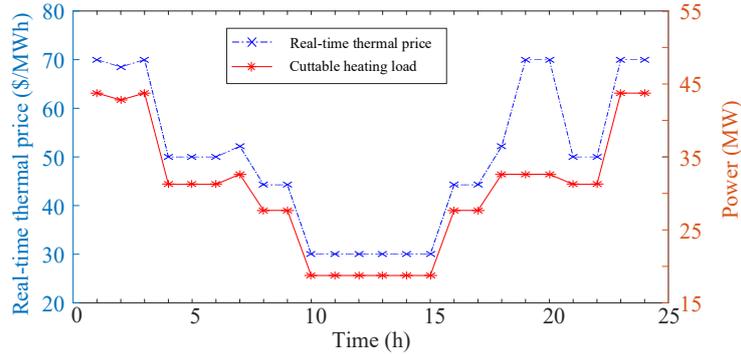

Fig.11. Real-time thermal price and the cuttable heating load

Fig. 11 illustrates that the real-time thermal prices are low in the periods 9:00-16:00, the cuttable heating loads are relatively light; in other periods, the real-time thermal prices are high, the cuttable heating loads are heavy. This phenomenon shows that RTP based IDR can guide users to actively adjust their consumption plan to minimize energy costs.

#### 5.1.4 Impact of penalty factors on cuttable heating load

To examine the impact of a penalty factor on the cuttable heating load, the cuttable heating load with different penalty factors is analyzed with the results shown in Fig. 12.

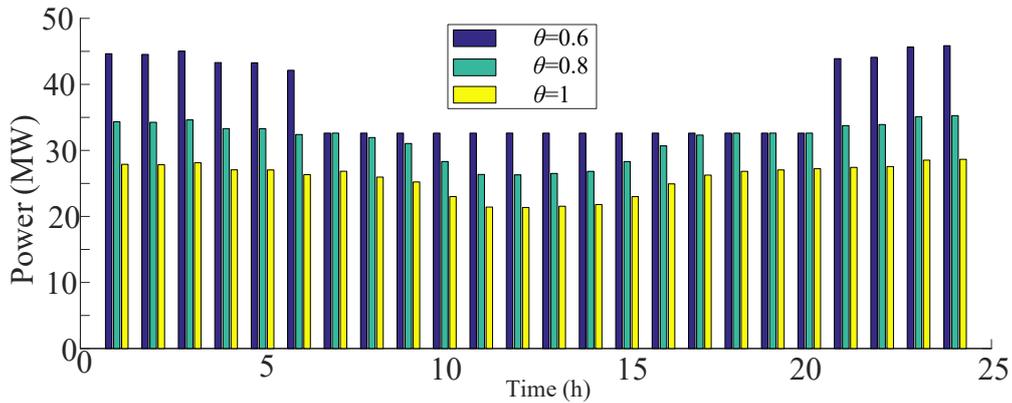

Fig. 12. Cuttable heating load in different penalty factors

As shown in Fig. 12, the cuttable heating load respectively reaches its maximum and minimum values when the penalty factor is taken as 0.6 and 1. For a same penalty factor, the cuttable heating loads during the nighttime periods (the thermal sensation is insensitive) are greater than other periods; and furthermore, this phenomenon is more obvious with a smaller penalty factor. Since the penalty factor depicts the users' demand for thermal comfort, the proper selection of this parameter plays an important role in the determination of cuttable heating loads.

#### 5.1.5 Analysis of confidence levels

It is of great significance to choose proper spinning reserves to balance the operational reliability and economy of the system. Thus, the reserve capacities under different confidence levels are given in Fig. 13.

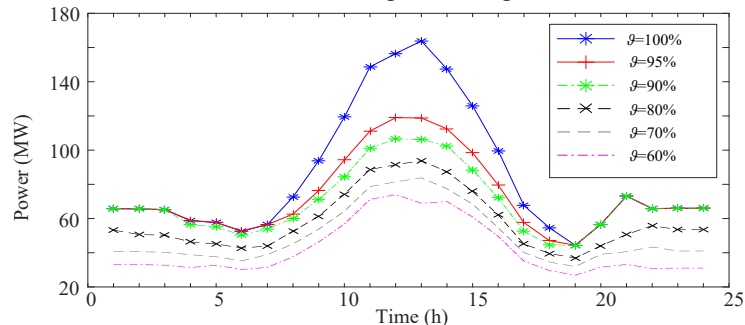

Fig. 13. Reserve capacities under different confidence levels

From Fig. 13, it can be found that as the confidence level increases, the required spinning reserve increases continuously, which improves the operational reliability of the system to a certain extent. But at the same time, the IES has to increase the outputs of the spinning reserve providers when doing so, so the operating cost of the system is increased. Therefore, the selection of an appropriate confidence level is crucial to realize the trade-off between reliability and economy.

### 5.1.6 Economy analysis

To reasonably evaluate the economy of the proposed approach, this study performs a comparative analysis of the IEO profits and the energy consumption costs of users in the four modes. The results are shown in Table 6.

Table 6. Profits of IEO and costs of users

| Mode | Profit of IEO ($) | Cost of users ($) |
| --- | --- | --- |
| Mode 1 | 1100941.45 | 1234584.68 |
| Mode 2 | 1101022.73 | 1234584.68 |
| **Mode 3** | **1069077.97** | **1195416.12** |
| Mode 4 | 1027523.82 | 1164671.31 |

Table 6 shows that due to the consideration of the DHN characteristics, the profit of the IEO in mode 2 is higher than that in mode 1, which verifies that considering the DHN characteristics can improve the IEO profit to a certain extent. Furthermore, through the economy comparison in modes 2, 3, and 4, it can be found that both the IEO profit and the energy cost of users in mode 3 lie halfway in between. This result demonstrates that the proposed approach manages to achieve the Stackelberg equilibrium of the benefits of IEO and users through the simulation of the interactive decision-making relationship between multi-stakeholders when they pursue their optimal objectives.

### 5.2 Case 2: a real integrated energy system in China

The Stackelberg game-based IES scheduling approach is applied to a real IES to examine its applicability for real-world applications. The real-world IES in Jilin Province, China, comprises 10 HEs, each HE supplies 200 households and the average heating area of each household is 100 m$^2$; the data of power and heating loads is taken from a real grid in Jilin Province, China [29]; the real-time price parameters are set according to the real electricity and thermal prices [29], i.e., $\mu_{max}$=0.087$/KWh, $\mu_{min}$=0.050$/KWh, $\mu_{av}$=0.0685$/KWh, $\gamma_{max}$=0.039$/KWh, $\gamma_{min}$=0.020$/KWh, $\gamma_{av}$=0.0295$/KWh; both the maximum power output of PV units and the WT rated power output are 300 KW; the parameters of supply pipelines are selected from reference [20]; TP units and CHP units with the total capacity of 5.17 MW are set to participate in the scheduling, and the main parameters are shown in Tables 7 and 8; the maximum and minimum capacities of the BESS are respectively 900 KW·h and 200 KW·h, the maximum charge and discharge power is 400 KW; other parameters are the same as the settings in Case 1.

Table 7. Main parameters of TP units

| $P_{max}^{TP}$ (KW) | $P_{min}^{TP}$ (KW) | $r_{u,i}^{TP}$ (KW/h) | $r_{d,i}^{TP}$ (KW/h) | $a_i^{TP}$ ($\times 10^{-3}$$/KW$^2$) | $b_i^{TP}$ ($\times 10^{-3}$$/KW) | $c_i^{TP}$ ($) | $\omega_i$ ($\times 10^{-3}$$/KW) |
| --- | --- | --- | --- | --- | --- | --- | --- |
| 500 | 50 | 175 | 175 | 0.012 | 17.82 | 10.150 | 13.7 |
| 250 | 25 | 125 | 125 | 0.069 | 26.24 | 31.670 | 13.2 |
| 250 | 25 | 125 | 125 | 0.028 | 37.69 | 17.940 | 13.2 |
| 420 | 50 | 210 | 210 | 0.010 | 12.88 | 6.778 | 14.2 |

Table 8. Main parameters of CHP units

| $P_{max}^{CHP}$ (KW) | $P_{min}^{CHP}$ (KW) | $H_{max}^{CHP}$ (KW) | $r_{u,i}^{CHP}$ (KW/h) | $r_{d,i}^{CHP}$ (KW/h) | $a_i^{CHP}$ ($\times 10^{-3}$$/KWh$^2$) | $b_i^{CHP}$ ($\times 10^{-3}$$/KWh) | $c_i^{CHP}$ ($/h) | $c_{V,i}$ | $c_{m,i}$ | $\varphi_i$ ($\times 10^{-3}$$/KW) |
| --- | --- | --- | --- | --- | --- | --- | --- | --- | --- | --- |
| 2000 | 375 | 2500 | 500 | 500 | 0.0044 | 13.29 | 39 | 0.15 | 0.75 | 16.2 |
| 2000 | 375 | 2500 | 500 | 500 | 0.0044 | 13.29 | 39 | 0.15 | 0.75 | 16.2 |

### 5.2.1 Analysis of renewable energy consumption

Renewable energy consumptions under modes 1-3 in the real system are analyzed with the results shown in Fig. 14.

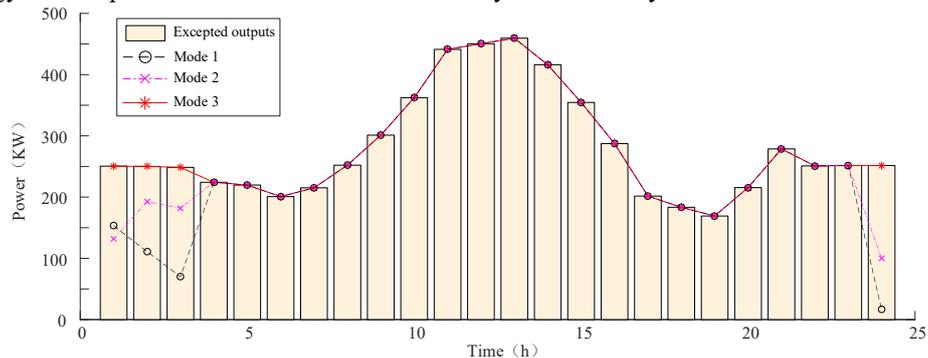

Fig. 14. Renewable consumptions under different modes in the real system

As shown in Fig. 14, the amounts of absorbed renewable energy under modes 1 and 3 are respectively the lowest and highest,

while the absorbed amount under mode 2 lies halfway in between. In particular, renewable energies are completely consumed under mode 3. The reason for the increasement of renewable consumptions is the DHN characteristics and IDR of the IES, which is in accordance with case 1.

### 5.2.2 Economy analysis

The IEO profits and the energy consumption costs of users in the real system are shown in Table 9.

Table 9. Profits of IEO and costs of users in the real system

| Mode | Profit of IEO ($) | Cost of users ($) |
|---|---|---|
| Mode 1 | 3936.49 | 8811.90 |
| Mode 2 | 3944.20 | 8811.90 |
| **Mode 3** | **3586.21** | **8448.76** |
| Mode 4 | 3197.47 | 8112.69 |

It can be seen from Table 9 that the proposed approach manages to achieve the Stackelberg equilibrium of the benefits of IEO and users in the real system. This result indicates that the presented method can also achieve a financial balance between multi-stakeholders for realistic systems, which demonstrates the applicability of the approach for real applications.

### 5.2.3 Analysis of calculation efficiency

To investigate the computational efficiency of the proposed solution method, the calculation times with varied confidence levels $\vartheta$ under different modes are shown in Table 10.

Table 10. Calculation times under different modes

| Operation modes | Calculation times (s) | | |
|---|---|---|---|
|  | $\vartheta=95\%$ | $\vartheta=90\%$ | $\vartheta=85\%$ |
| Mode 1 | 8.313 | 8.388 | 8.836 |
| Mode 2 | 8.626 | 8.275 | 8.573 |
| Mode 3 | 8.409 | 8.597 | 8.786 |
| Mode 4 | 8.152 | 8.954 | 8.448 |

Table 10 shows that the average calculation times in different conditions are around 8.5 seconds. Moreover, it can be anticipated that the calculation time can be further reduced by leveraging more advanced computer hardware configurations, more efficient programming language and parallel computing technology. Therefore, a conclusion can be drawn that the computational efficiency of the proposed solution method is able to satisfy the real-time requirements for IES scheduling in real applications.

## 6 Conclusion

To balance the integrated energy operator and users, this paper presents a Stackelberg game optimization framework for IES scheduling through coordination of renewable generations and IDR. Based on the simulation results, the following conclusions can be safely drawn:

(1) The proposed Stackelberg game-based modeling method of IES scheduling manages to achieve the Stackelberg equilibrium of the benefits of IEO and users through the leader-follower Stackelberg game.
(2) Due to the consideration of DHN characteristics and the flexible thermal comfort requirements of users in IDR, the approach is able to promote renewable consumptions and reduce the consumers' energy costs while maintaining consumers' thermal comfort within an acceptable range.
(3) The developed solution approach is capable of stably and fast obtaining the optimal solution of IEO and users under Stackelberg equilibrium.
(4) The test results on two cases verify the performances of adopted theoretical models and the solution methodology. Furthermore, the study on a real integrated energy system in China validate that the proposed approach is applicable for realistic applications.

Future work will focus on extending the proposed approach into the energy management of integrated electrical–heat–gas systems. Besides, it is assumed that the secondary heating network is ignored in this paper, while a more realistic scenario shall be considered.

**Acknowledgements**

This work is partly supported by the Natural Science Foundation of Jilin Province, China under Grant No. 2020122349JC.